\documentstyle[preprint,aps,epsf,floats,tighten]{revtex}

\begin{document}

\def\Bbar  {{\kern 0.18em\overline{\kern -0.18em B}}{}}
\def\Lbar{{\kern0.2em\overline{\kern-0.2em\Lambda\kern0.05em}%
    \kern-0.05em}{}}
\def\babar{\mbox{B\hspace{-0.4em} {\scriptsize A}\hspace{-0.4em} B\hspace{-0.4em}%
    {\scriptsize A\hspace{-0.1em}R}}}
\draft

\preprint{
\vbox{\hbox{JHU--TIPAC--98012}
      \hbox{hep-ph/9812217}
      \hbox{December 1998} }}

\title{Introduction to Hadronic $B$ Physics}
\author{Adam F.~Falk}
\address{Department of Physics and Astronomy, The Johns Hopkins
University\\ 3400 North Charles Street, Baltimore, Maryland
21218}
\maketitle
\thispagestyle{empty}
\setcounter{page}{0}
\begin{abstract}%
An overview of the theory of $B$ physics is given, with an emphasis on issues in the strong
interactions and hadronic physics.  This article is taken from an introductory chapter of {\sl The BaBar
Physics Book -- Physics at an Asymmetric B Factory}, SLAC Report SLAC-R-504.  It is written at the level
of a basic survey aimed at the experimental community.
\end{abstract}
\vskip3in
\centerline{Published as Chapter 2 of}
\smallskip
\centerline{{\sl The BaBar Physics Book -- Physics at an Asymmetric B Factory}}
\smallskip
\centerline{SLAC Report SLAC-R-504}

\newpage

\section{Introduction}

This review appeared originally as an introductory chapter of {\sl The BaBar Physics Book -- Physics at an
Asymmetric B Factory}, SLAC Report SLAC-R-504. As a whole, the book is a collaborative effort to
summarize and review the current (1998) state of the art in our understanding of $B$ physics, especially
as the theory will be applied to the analyses to be performed by the BaBar Collaboration.  The complete
text of {\sl The BaBar Physics Book} is available on the web site of the Stanford Linear Accelerator
Center, at {\tt http://www.slac.stanford.edu}.  However, this self-contained chapter
is being made available separately, because it might be of interest to an audience
somewhat broader than the experimental $B$ physics community toward which the rest of {\sl The
BaBar Physics Book} is aimed.  Its purpose is to provide an overview of, and basic
introduction to, theoretical techniques essential to the study of $B$ mesons and their
decays.  Of course, there is an enormous literature on this subject, and it is not the
goal here to review it all.  Rather, this introduction provides a general context within
which to place the theoretical treatments of specific processes discussed elsewhere in
{\sl The BaBar Physics Book}.  The focus here is on  general issues of philosophy and approach,
and the goal is to be pedagogical. No attempt is made to provide a comprehensive review of
the field, nor to include references to the voluminous literature of individual
contributions on each topic.  Instead, at the end are some suggestions for further
reading.  For a fully referenced review of theoretical $B$ physics, or at least of those
parts of direct relevance to BaBar, the reader should see {\sl The BaBar Physics Book}
itself. 

A key factor in the experimental interest in $b$ physics is the potential insight it
affords into physics at very short distances.  In particular, it is hoped that the high
precision study of phenomena such as $CP$ violation, rare decays, and flavor changing
processes will provide precious insights into new interactions associated with the flavor
sector of whatever theory lies beyond the Standard Model.  However, in order for this
information to become  available, it is necessary to confront the fact that the $b$
quarks, which are the  ultimate objects of study, are bound by strong dynamics into color
neutral hadrons.  While understood in principle, the nonperturbative nature of these bound
states makes problematic the extraction of precision information about physics at  higher
energies from even the most clever and precise experiments on $B$ mesons.   To explore new
physics effects one faces a daunting theoretical challenge to untangle  them from the
effects of nonperturbative QCD.

This is not a problem which has been solved in its entirety, nor is it likely  ever to
be.  Rather, what is available is a variety of theoretical approaches and techniques,
appropriate to a variety of specific problems and with varying levels of reliability. 
There are a few situations in which one can do analyses which are rigorous and predictive,
and many in which what can be said  is more imprecise and model dependent.  The result is
an interesting interplay between theory and experiment, where one often cannot measure
what one can compute reliably, nor compute reliably what one can measure.  In the search
for quantities which can be both predicted and measured, one must be creative and flexible
in the choice of theoretical techniques.  While approaches which are based directly on
QCD, and which allow for quantitative error estimates,  are clearly to be preferred, more
model-dependent methods are often all that are available and thus have an important role
to play as well.

The theoretical methods  discussed here fall roughly  into three categories.
First, there are effective field theories such as the Heavy  Quark Expansion  (HQE) and
Chiral Perturbation Theory (ChPT).  Effective field theories  derive their  predictive
power by exploiting systematically a small expansion parameter.   For  nonperturbative
QCD, this parameter cannot be the strong coupling constant 
$\alpha_s$; instead, it is a ratio of mass scales obtained by considering a particular 
limit or  special kinematics.  Second, there are the approaches of lattice QCD and  QCD
sum rules, which are based on QCD but do not exploit a large separation of scales.   While
in  principle these techniques are rigorous, they suffer in their current  practical 
implementations from a degree of uncontrolled model dependence.  In the  case of the 
lattice, this problem will improve with the availability of ever more powerful 
computers.  Third, there are quark models, which do not purport to be  derived from QCD. 
Instead, in using models one introduces some new degrees  of freedom and  interactions 
which, it is hoped, capture or mimic some behavior  of the true theory.  The  advantage
of  models is their flexibility, since a model  may be tuned to particular  processes or 
hadronic states.  The disadvantage is that
 models are intrinsically  {\it ad hoc,} and  it is difficult to assess their reliability. 
 For this reason, one should use them  only when no better options are available. 

Effective field theories are based on the idea that in a given process, only  certain
degrees of freedom may be important for understanding the physics.  In  particular,  it is
often the case that kinematical considerations which restrict the  momenta of  external
particles effectively restrict the momenta of virtual particles  as well.   Thus it is
sensible to remove from the theory intermediate states of high  virtuality.  Their absence
may be compensated by introducing new ``effective''  interactions between  the degrees of
freedom which remain.  Effective field theories  are often  constructed  using the
technique of the operator product expansion,  which provides an  elegant and  general
conceptual framework.

Both the HQE and ChPT are effective field theories which are derived from  formal  limits
of QCD in which the theory exhibits new and useful symmetries.  In the  case of the HQE,
the limit is $m_b, m_c\to\infty$, where a ``spin-flavor''  symmetry yields a variety of
predictions for heavy hadron spectroscopy and  semileptonic decays.  For ChPT, the limit
is $m_u, m_d, m_s\to0$, which leads  to  exact predictions for the emission and absorption
of soft pions.  In both  cases, the quark masses are large or small compared to the scale
of nonperturbative  QCD,  typically hundreds of MeV.  What makes an effective field theory
powerful is  that the deviations from the limiting behavior may be organized in a
systematic  expansion  in a small parameter.  Hence one can both improve the accuracy of
an analysis  and  derive quantitative error estimates.  An effective field theory is
predictive  precisely because it is under perturbative control.

While the HQE and ChPT are powerful tools where they may be applied, their  use is 
restricted to a small number of processes involving certain initial and final  states. 
Unfortunately, the HQE and ChPT have nothing to say about the vast  majority of processes
and quantities available for experimental study at a $B$  Factory.  Similar considerations
affect lattice QCD.  Because of both  computational  and theoretical limitations, reliable
lattice predictions are confined  largely to  spectroscopy and matrix elements with
restricted kinematics.  QCD sum rules,  also for technical reasons, may only be used in
limited circumstances.

Thus  a serious problem remains, namely that many quantities of  experimental  and
phenomenological importance cannot be analyzed by methods which are  systematic and well
understood.  For inclusive weak decays, some exclusive semileptonic  decays,  and some
static properties,  effective field  theories  or the lattice give controlled theoretical
predictions.  But for the  description of exclusive hadronic weak decays,  most  exclusive
semileptonic decays, strong decays, fragmentation, and many other  interesting  aspects of
$B$ physics, only a variety of model dependent  approaches are available.  While no model
is ``correct'', some models are better than  others.   A successful model should be
motivated by some physical picture, should  reproduce  much more data than there are input
parameters, and should behave correctly in  appropriate limits, such as obeying heavy
quark symmetry as $m_b\to\infty$.   It  will not be possible here to discuss or
even enumerate all of the  models  which are used in $B$ physics, but it is generally true
that every model  ought to be  judged by criteria such as these.

Because there is no single theoretical framework which suffices for all of $B$  physics,
it is often necessary to utilize a variety of methods in one  theoretical  analysis. 
Usually, this is desirable, as a combination of complementary  approaches  can lead to
conclusions which are much more robust.  But at the same time,  one must be  careful to be
consistent in the use and definition of theoretical concepts and  quantities, and
particularly in their translation from one context to  another.   Otherwise one is led
easily to error and confusion.

An excellent illustration of how problems can arise is given by the  definition of the 
heavy quark mass.  Clearly there is {\it something\/} which is meant by  ``the $b$ mass'',
because to say that the $b$ quark is heavy is to say that the  parameter $m_b$  is large 
compared to $\Lambda_{\rm QCD}$.  Whatever the $b$ mass is, it is presumably  somewhere 
close to 4 or 5~GeV.  But the situation becomes more complicated when one  tries to  pin
down $m_b$ more precisely than that.

On the one hand, it is known that the $b$ quark acquires its mass from its  coupling,  of
strength $\lambda_b$, to the ``Higgs vacuum expectation value'' $v$, so 
$m_b=\lambda_b v$.  The quark mass which is directly related to this coupling  is  known
as the ``current mass'' or ``short distance mass.''  Its value depends  on the 
renormalization scheme, such as $\overline{\rm MS}$, which is used to define  the 
theory.  In perturbation theory, there is also a pole in the $b$ quark  propagator,  the
position of which corresponds to the rest energy of a freely propagating 
$b$  quark.  This ``pole mass'' is closer to an intuitive notion of an invariant, 
relativistic mass.  Unfortunately, because of confinement, a freely  propagating  $b$
quark cannot actually exist, and the pole mass is not defined  nonperturbatively.  In
fact, even within perturbation theory the pole mass is ill  behaved and  can only  be
defined to a fixed finite order $\alpha_s^n$.  Hence  there is really a  family  of pole
masses, namely the ``1-loop pole mass'', the  ``2-loop pole mass'',  and so  on, none more
``accurate'' or intuitively accessible than another.  There are  also ``Wilsonian''
running masses $m_b(\mu)$, which are defined with  additional  subtractions in the
infrared.

An analogous variety of $b$ quark masses is defined in lattice calculations.   While it is
typically understood how these lattice $b$ masses are related to  each other, relating
them to pole or current masses defined in continuum QCD  can be problematic.  For example,
lattice field theory, both perturbative and  nonperturbative, is regulated and subtracted
differently from field theory in  the continuum, and the relationship between the various
schemes often is not  straightforward.  Similar ambiguities can affect the $b$ quark
masses which  appear in QCD sum rules.  Finally, there are the many quark masses
introduced  in models, which are free parameters with {\it no\/} rigorous relationship 
either  to each other or to masses defined in QCD.  A typical example is the 
``constituent  quark mass'' of the nonrelativistic quark model.  No matter how  precisely
one  fits the constituent quark mass to data, it can never be used as an input  into a 
lattice or HQE calculation.  The most that can be said is that all of these  various 
masses probably are within several hundred MeV of each other.   

It is important to understand that there is no more precise way to unify  these many 
masses into a single universal quantity.  The ambiguity in $m_b$ is  unimportant, so long
as its definition is {\it consistent\/} within a given analysis, and  ultimately  one
predicts measurable quantities in terms of other measurable quantities.  The
problem is that it is difficult to make an $m_b$ defined on the lattice  consistent  with
one defined in the continuum, and impossible to make a model dependent 
$m_b$  consistent with either.  Hence there can be limits {\it in principle\/} to the 
accuracy which one can obtain when a variety of methods are combined in a  single analysis.

We now turn to elementary introductions to the most important theoretical techniques in
$B$ physics.  After a general discussion of operator  product expansions and effective
field theories,  Heavy Quark  Effective Theory and Chiral Perturbation Theory are
introduced.  The next two sections contain  discussions of lattice QCD and QCD sum rules,
followed by a  brief discussion of quark models.  None of these ideas
will be developed  in much  depth.  Rather, they are intended to serve as a background to
the variety of detailed theoretical analyses which are presented in {\sl The BaBar Physics
Book.}

\section{The Operator Product Expansion}
\label{Chap2:secAA}

\subsection{General Considerations}
\label{Chap2:ssecAA}

A central observation which underlies much of the theoretical study of
$B$ mesons is that physics at a wide variety of distance (or momentum) scales is typically
relevant in a given process.  At the same time, the physics at different scales must often
be analyzed with different theoretical approaches.  Hence it is crucial to have a tool
which enables one to identify the physics at a given scale and to separate it out
explicitly.  Such a tool is the operator product expansion, used in  conjunction with the
renormalization group.  Here
 a general discussion of its application is given.

\begin{figure}
\epsfxsize=6cm
\hfil\epsfbox{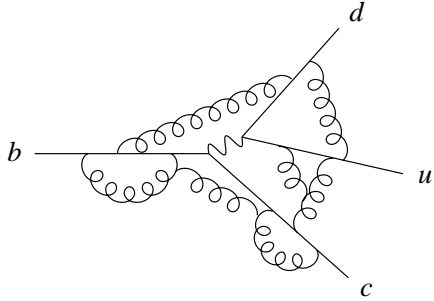}\hfill
\caption{The nonleptonic decay of a $b$ quark.}
\label{Chap2:Chap2fig1}
\end{figure}

Consider the Feynman diagram shown in Fig.~\ref{Chap2:Chap2fig1}, in which a 
$b$ quark decays nonleptonically.  The virtual quarks and gauge bosons have virtualities
$\mu$ which vary widely, from $\Lambda_{\rm QCD}$ to $M_W$ and higher.  Roughly speaking,
these virtualities can be classified into a variety of energy regimes: (i) $\mu\gg M_W$;
(ii) $M_W\gg\mu\gg m_b$; (iii) $m_b\gg\mu\gg\Lambda_{\rm QCD}$; (iv)
$\mu\approx\Lambda_{\rm QCD}$.  Each of these momenta corresponds to a different distance
scale; by the uncertainty principle, a particle of virtuality $\mu$ can propagate a
distance $x\approx 1/\mu$ before being reabsorbed.  At a given resolution $\Delta x$, only
some of these virtual particles can be distinguished, namely those that propagate a
distance $x>\Delta x$.  For example, if $\Delta x>1/M_W$, then the virtual $W$ cannot be
seen, and the process whereby it is exchanged would appear as a point interaction.  By the
same token, as
$\Delta x$ increases toward $1/\Lambda_{\rm QCD}$, fewer and fewer of the virtual gluons
can be seen explicitly.  Finally, for
$\mu\approx\Lambda_{\rm QCD}$, it is probably not appropriate to speak of virtual gluons
at all, because at such low momentum scales QCD becomes strongly interacting and a
perturbation series in terms of individual gluons is inadequate.

It is useful to organize the computation of a diagram such as is shown in
Fig.~\ref{Chap2:Chap2fig1} in terms of the  virtuality of the exchanged particles.  This
is important both conceptually and practically.  First, it is often the case that a
distinct set of approximations and approaches is useful at each distance scale, and one 
would like to be able to apply specific theoretical techniques at the scale at which they
are appropriate.  Second, Feynman diagrams in which two distinct scales $\mu_1\gg\mu_2$
appear together can lead to logarithmic corrections of the form
$\alpha_s\ln(\mu_1/\mu_2)$, which for $\ln(\mu_1/\mu_2)\sim1/\alpha_s$ can spoil the
perturbative expansion.  A proper separation of scales will include a resummation of such
terms.

\subsection{Example I:  Weak $b$ Decays}
\label{Chap2:ssecAB}

As an example, consider the weak decay of a $b$ quark, $b\to c\bar u d$, which is mediated
by the decay of a virtual $W$ boson.  Viewed with resolution $\Delta x<1/M_W$, the decay
amplitude involves an explicit
$W$ propagator and is proportional to
\begin{equation}
\label{Chap2:eq:matel1}
  \bar c\gamma^\mu(1-\gamma^5)b\,\bar d\gamma_\mu(1-\gamma^5)u
  \times{(ig_2)^2/4\over p^2-M_W^2}\,,
\end{equation} where $p^\mu$ is the momentum of the virtual $W$.  Since $m_b\ll M_W$, the
kinematics constrains $p^2\ll M_W^2$, so the virtuality of the $W$ is of order $M_W$, and
it travels a distance of order $1/M_W$ before decaying.  Viewed with a lower resolution,
$\Delta x>1/M_W$, the process $b\to c\bar u d$ appears to be a local interaction, with four
fermions interacting via a potential which is a $\delta$ function where the four particles
coincide.  This can be seen by making a Taylor expansion of the amplitude in powers of
$p^2/M_W^2$,
\begin{equation}
  \bar c\gamma^\mu(1-\gamma^5)b\,\bar d\gamma_\mu(1-\gamma^5)u
  \times{g_2^2\over8M_W^2}\left[1+{p^2\over M_W^2}
  +{p^4\over M_W^4}+\dots\right]\,.
\end{equation} The coefficient of the first term is just the usual Fermi decay constant,
$G_F/\sqrt2$.  The higher order terms correspond to local operators of higher mass
dimension.  In the sense of a Taylor expansion, the momentum-dependent matrix element
(\ref{Chap2:eq:matel1}), which involves the propagation of a $W$ boson between {\it two\/}
spacetime points, is identical to the matrix element of the following infinite sum of
local operators:
\begin{equation}
\label{Chap2:eq:nonlepeff}
  {G_F\over\sqrt2}\,\bar c\gamma^\mu(1-\gamma^5)b
  \left[1+{(i\partial)^2\over M_W^2}+{(i\partial)^4\over M_W^4}
  +\dots\right]
  \bar d\gamma_\mu(1-\gamma^5)u\,,
\end{equation} where the derivatives act on the entire current on the right.  This
expansion of the nonlocal product of currents in terms of local operators,  sometimes
known as an {\it operator product expansion,} is valid so long as  
$p^2\ll M_W^2$.  For $B$ decays, the external kinematics requires 
$p^2\le m_b^2$,  so this condition is well satisfied.  In this regime, one may consider a
nonrenormalizable {\it effective field theory,} with interactions of dimension six and
above.  The construction of such a low energy effective theory is  also known as {\it
matching.}  As it is nonrenormalizable, the effective theory is defined (by construction)
only up to a cutoff, in this case $M_W$.  The cutoff is explicitly the mass of a particle
which has been removed from the theory, or {\it integrated out.}  If one considers
processes  in which one is restricted kinematically to momenta well below the cutoff, the
nonrenormalizability of the theory poses no technical problems.  Although the coefficients
of operators of dimension greater than six require counterterms  in the effective theory
(which may be unknown in strongly interacting  theories),   their matrix elements are
suppressed by  powers of
$p^2/M_W^2$.  To a {\it given order\/} in $p^2/M_W^2$, the theory is well-defined and
predictive.

From a modern point of view, in fact, such nonrenormalizable effective theories are
actually preferable to renormalizable theories, because the nonrenormalizable terms
contain information about the energy  scale at  which the theory ceases to apply.  By
contrast, renormalizable  theories contain no such explicit clues about their region of
validity.

In principle, it is possible to include effects beyond leading order in
$p^2/M_W^2$ in the effective theory, but in practice, this is usually quite complicated
and rarely worth the effort.  Almost always, the operator product expansion is truncated
at dimension six, leaving only the four-fermion contact term.  Corrections to this
approximation are of order
$m_b^2/M_W^2\sim10^{-3}$.

\subsection{Radiative Corrections}

\begin{figure}
\epsfxsize=6cm
\hfil\epsfbox{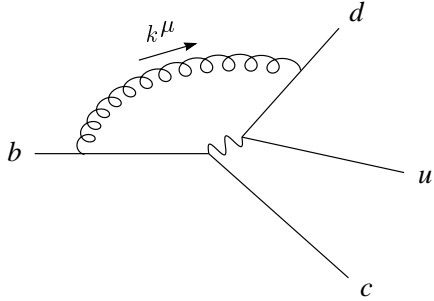}\hfill
\caption{The nonleptonic decay of a $b$ quark at one loop.}
\label{Chap2:Chap2fig2}
\end{figure}

At tree level, the effective theory is constructed simply by integrating out the $W$
boson, because this is the only particle in a tree level diagram which is off-shell by
order $M_W^2$.  When radiative corrections are included, gluons and light quarks can also
be off-shell by this order.  Consider the one-loop diagram shown in
Fig.~\ref{Chap2:Chap2fig2}.   The components of the loop momentum $k^\mu$ are allowed to
take all values in the loop integral.  However, the integrand is cut off both in the
ultraviolet and in the infrared.  For $k>M_W$, it scales as $d^4k/k^6$, which is
convergent as $k\to\infty$.  For $k<m_b$, it scales as $d^4k/k^3m_bM_W^2$, which is
convergent as $k\to0$.  In between, all momenta in the range
$m_b<k<M_W$ contribute to the integral with roughly equivalent weight. 

As a consequence, there is potentially a radiative correction of order
$\alpha_s\ln(M_W/m_b)$.  Even if $\alpha_s(\mu)$ is evaluated at the high scale $\mu=M_W$,
such a term is not small in the limit $M_W\to\infty$.  At
$n$ loops, there is potentially a term of order
$\alpha_s^n\ln^n(M_W/m_b)$.  For $\alpha_s\ln(M_W/m_b)\sim1$, these terms need to be
resummed for the perturbation series to be predictive.  The technique for performing such
a resummation is the {\it renormalization group.}

The renormalization group exploits the fact that in the effective theory, operators such
as 
\begin{equation}
\label{Chap2:eqI}
  O_I=\bar c_i\gamma^\mu(1-\gamma^5)b^i\,
  \bar d_j\gamma_\mu(1-\gamma^5)u^j
\end{equation}   receive radiative corrections and must be subtracted and renormalized. 
(Here the color indices $i$ and 
$j$ are explicit.)  In dimensional regularization, this means that they acquire, in 
general,  a dependence on the renormalization scale $\mu$.  Because physical  predictions
are of necessity independent of $\mu$, in the renormalized effective theory it must be the
case that the operators are multiplied by coefficients with a dependence on $\mu$ which
compensates that of the operators.  It is also possible for operators to mix under
renormalization, so the set of operators induced at tree level may be enlarged once
radiative corrections are included.  In the present example, a second operator with
different color structure,
\begin{equation}
\label{Chap2:eqII}
  O_{II}=\bar c_i\gamma^\mu(1-\gamma^5)b^j\,
  \bar d_j\gamma_\mu(1-\gamma^5)u^i\,,
\end{equation} is induced at one loop. The interaction Hamiltonian of the effective theory
is then
\begin{equation}
\label{Chap2:eq:Heff}
  {\cal H}_{\rm eff}=C_I(\mu)O_I(\mu)+C_{II}(\mu)O_{II}(\mu)\,,
\end{equation} and it satisfies the differential equation
\begin{equation}
  \mu{d\over d\mu}{\cal H}_{\rm eff}=0\,.
\end{equation} By computing the dependence on $\mu$ of the operators $O_i(\mu)$, one can
deduce the $\mu$-dependence of the {\it Wilson coefficients\/} 
$C_i(\mu)$.   In this case, a simple calculation yields
\begin{equation}
\label{Chap2:eq:lowil}
  C_{I,II}(\mu)=
  {1\over2}\left[\left({\alpha_s(M_W)\over\alpha_s(\mu)}\right)^{6/23}
  \pm\left({\alpha_s(M_W)\over\alpha_s(\mu)}\right)^{-12/23}\right]\,.
\end{equation} For $\mu=m_b$, these expressions resum all large logarithms proportional to
$\alpha_s^n\ln^n(M_W/m_b)$.

The decays which are observed involve physical hadrons, not asymptotic quark states.  For
example, this nonleptonic $b$ decay can be realized in the channels $B\to D\pi$, $B\to
D^*\pi\pi$, and so on.  The computation of partial decay rates for such processes requires
the analysis of hadronic matrix  elements such as 
\begin{equation}
  \langle D\pi|\,\bar c\gamma^\mu(1-\gamma^5)b\,
  \bar u\gamma_\mu(1-\gamma^5)d\,|\bar b\rangle\,.
\end{equation} Such matrix elements involve nonperturbative QCD and are extremely
difficult to compute from first principles.  However, they have no intrinsic dependence on
large mass scales such as $M_W$.  Because of this, they should naturally be evaluated at a
renormalization scale $\mu\ll M_W$, in which case large  logarithms 
$\ln(M_W/m_b)$ will not arise in the matrix elements.    By choosing such a  low scale in
the effective theory (\ref{Chap2:eq:Heff}),  all such terms are resummed into the
coefficient functions $C_i(m_b)$.  As promised, the physics at scales near $M_W$ has been
separated from the physics at scales near
$m_b$, with the renormalization group used to resum the large logarithms which connect
them. In fact, nonperturbative hadronic matrix elements are usually evaluated at an even
lower scale $\mu\approx\Lambda_{\rm QCD}\ll m_b$, explicitly resumming all perturbative
QCD corrections.

\subsection{Example II:  Penguins and Box Diagrams}
\label{Chap2:ssecAC}

\begin{figure}
\epsfxsize=12cm
\hfil\epsfbox{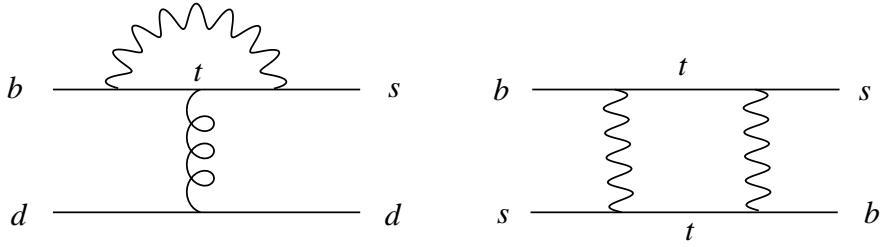}\hfill
\caption{Penguin diagram (left) and box diagram (right).}
\label{Chap2:Chap2fig3}
\end{figure}

In the previous example of nonleptonic decays, the operator $O_I$ appeared  when the
matching at tree level was performed.  It is also  possible to find new operators in the
effective theory which appear only when the  matching is performed at one loop.  The most
common such operators are those which arise from penguin and box diagrams, such as those
shown in  Fig.~\ref{Chap2:Chap2fig3}.  These  diagrams are important in $b$ physics
typically because they lead to  flavor-changing interactions at low energies which are
suppressed at tree level in the Standard Model.  Hence the transitions mediated by these
operators are potentially a sensitive probe of new physics.

Both penguins and box diagrams can lead, via the operator product expansion, to four-quark
operators with new flavor or color structure, such as 
\begin{equation}
  \bar s\gamma^\mu T^a b\,\bar d\gamma_\mu T^ad\,,
\end{equation} which mediates nonleptonic $B$ decay, or
\begin{equation}
  \bar b\gamma^\mu\gamma^5 d\, \bar b\gamma_\mu\gamma^5 d\,,
\end{equation} which is responsible for $B^0$--$\Bbar^0$ mixing. Penguins can also lead to new
flavor-changing magnetic interactions, such as 
\begin{equation}
  \bar s\sigma^{\mu\nu}T^ab\,G^a_{\mu\nu}\,,
\end{equation} when the $d$ quark line in Fig.~\ref{Chap2:Chap2fig3} is removed.  The
gluon  could also be  replaced by a photon or a $Z$ boson.  From the point of view of the
low energy effective theory, it is unimportant that these operators  arise at one loop at
high energy.  They can mix with four-fermion operators induced at tree level, insofar as
such mixing is allowed by the flavor and Lorentz symmetries of the effective theory.  In
fact, the renormalization of penguin-induced operators can be quite complicated, due to
the nature of their flavor structure; two loop calculations may be required to resum the
leading logarithms $\alpha_s^n\ln^n(M_W/m_b)$.

\subsection{Summary}
\label{Chap2:ssecAD}

Low-energy effective theories are constructed using the operator product expansion and the
renormalization group.   This procedure implements an important separation of scales,
isolating the physics which involves  virtualities $\mu\gg m_b$ and accounting for it
systematically in a double expansion in  powers of $\alpha_s$ and $m_b/M_W$, where $M_W$
is the matching scale at which heavy particles are integrated out of the theory. This
procedure may be generalized to integrate out heavy particles of many different kinds.

This analysis explicitly does not address those parts of a process which are  dominated 
by low momenta, which will typically be more difficult to deal with.  By  breaking  the
problem up according to momentum scale, one may compute systematically in  perturbation
theory where it is possible to do so.  However, the accuracy  obtained  in this part of
the calculation is useful only if one can also account for  physics  at lower energy.  The
chief limitation on the accuracy of most theoretical  calculations in $b$ physics is, in
fact, from these lower energy effects.

\section{The Heavy Quark Expansion}
\label{Chap2:secBB}

\subsection{Separation of Scales}
\label{Chap2:ssecBA}

This section considers physics characterized by virtualities $\mu\approx m_b$ and  below. 
The previous section discussed how physics at higher scales is accounted  for in QCD
perturbation theory, because at high energies $\alpha_s(\mu)/\pi\ll 1$.  Although $m_b\ll
M_W$, at this ``low'' energy it is still the case that
$\alpha_s(m_b)/\pi\approx0.1\ll 1$, and $\Lambda_{\rm QCD}/m_b\sim0.1\ll 1$.  Hence one
seeks a technique analogous to the operator product expansion by which to exploit the
presence of such small parameters.

The status of the $b$ quark in a $B$ meson is different from that of a virtual $W$ in a
weak decay, because the $b$ is real, not virtual, and the $B$ carries nonzero $b$-number
which persists in the asymptotic state.  Hence it is not appropriate to integrate out the
$b$ in the same sense as the $W$ was integrated  out, removing it from the theory
entirely.  Rather,  when bound into a hadron with light degrees of freedom of typical
energies
$E\sim\Lambda_{\rm QCD}$, the $b$ makes excursions from its mass shell by virtualities
only of order $\Lambda_{\rm QCD}$.  What can  be integrated out is not the $b$ itself, but
rather those parts of the $b$ field which take it far off shell.  The result will be an
effective theory of a  static $b$ quark, in its rest frame.  

Processes with hard virtual gluons, which drive the $b$ far off shell, will  lead to
perturbative corrections in the effective theory of order $\alpha_s(m_b)$.  They may be
included as before.  In addition, {\it power corrections} appear, analogous to the higher
order operators appear in Eq.~(\ref{Chap2:eq:nonlepeff}).  In this case, it will be
necessary to  include the leading higher-dimension operators to achieve results of the
desired accuracy.  These power corrections will lead to terms of order  $(\Lambda_{\rm
QCD}/m_b)^n$.  The appearance of the scale $\Lambda_{\rm QCD}$ serves as a reminder that
these corrections involve nonperturbative physics, and will typically not be calculable
from first principles.  Instead, the inclusion of power corrections will require the
introduction of new phenomenological parameters, whose values are controlled by
nonperturbative QCD.  These parameters have precise field-theoretic definitions, and they
will be introduced in a systematic  manner.  Their appearance will not spoil the inherent
predictability of the theory,  although in practice they will increase the number of
quantities which must be  determined from experiment before accurate predictions can be
made.

Finally, for some applications (notably the analysis of exclusive semileptonic
$B$ decays), it will be useful to treat the $c$ quark as heavy, that is, to perform an
expansion also in powers of $\Lambda_{\rm QCD}/m_c$.  In this case, clearly, the leading
power corrections will be important and will have to be well understood for the theory to
be predictive.

\subsection{Heavy Quark Symmetry}
\label{Chap2:ssecBB}

Let us, for generality, consider a hadron $H_Q$ composed of a heavy quark $Q$  and ``light
degrees of freedom'' consisting of quarks, antiquarks and gluons, in the limit
$m_Q\to\infty$.  The Compton wavelength of the heavy quark scales as the inverse of the
heavy quark mass, $\lambda_Q\sim1/m_Q$.  The light degrees of freedom, by contrast, are
characterized by momenta of order
$\Lambda_{\rm QCD}$, corresponding to wavelengths 
$\lambda_\ell\sim1/\Lambda_{\rm QCD}$.  Since $\lambda_\ell\gg\lambda_Q$, the light
degrees of freedom cannot resolve features of the heavy quark other than its conserved
gauge quantum numbers.  In particular, they cannot probe the actual {\it value\/} of
$\lambda_Q$.  Although the structure of the hadron
$H_Q$ is determined by nonperturbative strong interactions, the typical momenta exchanged
by the light degrees of freedom with each other and with the heavy quark are of order
$\Lambda_{\rm QCD}\ll m_Q$, against which the heavy quark 
$Q$ does not recoil.  In this limit, $Q$ acts as a static source of electric and
chromoelectric field.

There is an immediate implication for the spectroscopy of heavy hadrons.  Since the
interaction of the light degrees of freedom with the heavy quark is independent of $m_Q$,
then so is the spectrum of excitations.  It is these excitations which determine the
spectrum of heavy hadrons $H_Q$.  Since the  {\it splittings\/} $\Delta_i\sim\Lambda_{\rm
QCD}$ between the various hadrons 
$H_Q^i$ are entirely due to the properties of the light degrees of freedom,  they are
independent of $Q$ and, in the limit $m_Q\to\infty$, do not scale  with $m_Q$. For
example, if $m_b,m_c\gg\Lambda_{\rm QCD}$, then the light degrees of freedom are in
exactly the same state in the mesons $B_i$ and $D_i$, for a given $i$.  The offset
$B_i-D_i=m_b-m_c$ is just the difference between  the heavy quark masses.  By no means
does the relationship between the spectra  rely on an approximation $m_b\approx m_c$.

This picture is enriched by recalling that the heavy quarks and light degrees  of freedom
also carry angular momentum.  The heavy quark has spin quantum number
$S_Q={1\over2}$, which leads to a chromomagnetic moment 
$\mu_Q\propto g/2m_Q$.  Note that $\mu_Q\to0$ as $m_Q\to\infty$, and the interaction
between the spin of the heavy quark and the light degrees of freedom is suppressed.  Hence
the light degrees of freedom are insensitive to $S_Q$; their state is independent of
whether $S_Q^z={1\over2}$ or $S_Q^z=-{1\over2}$.  Thus each of the energy  levels $B_i$
and $D_i$ is actually doubled, one state for each value of 
$S_Q^z$.  In summary, the light degrees of freedom in a heavy hadron are the  same when
combined with any of the four heavy quark states:
\begin{equation}
  b(\uparrow)\,,\quad b(\downarrow);\quad 
  c(\uparrow)\,,\quad c(\downarrow)\,.
\end{equation} The result is an $SU(4)$ symmetry which leads to nonperturbative relations 
between  physical quantities.

Suppose the light degrees of freedom have total angular momentum $J_\ell$,  which is 
integral for baryons and half-integral for mesons.  When combined with the   heavy  quark
spin $S_Q={1\over2}$,  physical hadron states can be produced with total angular  momentum
\begin{equation}
  J=\left| J_\ell\pm\textstyle{{1\over2}} \right|\,.
\end{equation} If $J_\ell\ne0$, then these are two degenerate states.  For example, the 
lightest heavy mesons have $J_\ell=\textstyle{{1\over2}}$, leading to a doublet with 
$J=0$ and $J=1$.  When effects of order $1/m_Q$ are included, the chromomagnetic
interactions split the states of given $J_\ell$ but different $J$.  This ``hyperfine''
splitting is not calculable perturbatively, but it is  proportional to the heavy quark
magnetic moment $\mu_Q$.  Since $\mu_Q\propto1/m_Q$, one can  construct a relation which
is a nonperturbative prediction of heavy quark symmetry,
\begin{equation}
\label{Chap2:eq:spinsplitting}
  m_{B^*}^2-m_B^2=m_{D^*}^2-m_D^2\,.
\end{equation} Experimentally, $m_{B^*}^2-m_B^2=0.49\,{\rm GeV}^2$ and
$m_{D^*}^2-m_D^2=0.55\,{\rm GeV}^2$.  The correction to this prediction is of  order
$\Lambda^3_{\rm QCD}(1/m_c-1/m_b)\sim0.1\,{\rm GeV}^2$, so it works  about  as well as one
should expect.  Note that the relation (\ref{Chap2:eq:spinsplitting})
 involves not only the heavy quark  symmetry, but also the systematic inclusion of the
leading symmetry violating effects.

So far, heavy quark symmetry has been formulated for hadrons in their rest frame.  One can
easily boost to a frame in which the hadrons have arbitrary four-velocity
$v^\mu=\gamma(1,\vec v)$. The symmetry will then relate hadrons $H_b(v)$ and $H_c(v)$ with
the same velocity but with different  momenta.  This distinguishes heavy quark symmetry
from ordinary symmetries of QCD, which relate states of the same momentum.  It will often
be convenient to label heavy hadrons explicitly by their velocity: $B(v)$, $B^*(v)$, and
so on.

\subsection{Heavy Quark Effective Theory}
\label{Chap2:ssecBC}

It is quite useful to make heavy quark symmetry manifest within QCD by taking  the limit
$m_b\to\infty$ of the QCD Lagrangian.  This is done by making the  dependence of all
quantities on $m_b$ explicit, and then developing the Lagrangian  in a series in inverse
powers of $m_b$.  The idea is to write the Lagrangian in a  form in which the action of 
the heavy quark symmetries is well-defined at each order in the expansion, so  the effect
of symmetry breaking corrections can be studied in a systematic way.  The resulting
Lagrangian is known as the Heavy Quark Effective Theory (HQET).  The HQET is similar to an
effective theory which results from an operator product expansion, in the sense that the
only virtualities $p$ which are allowed  satisfy $p\ll m_b$, with effects of greater
virtuality absorbed into the coefficients  of higher dimension operators. The difference
is that in this case, the heavy $b$ quark is not explicitly removed from the effective
theory.

In the heavy quark limit, the velocity $v^\mu$ of the $b$ quark is conserved.  Thus one
may write its four-momentum in the form $p_b^\mu=m_b v^\mu+k^\mu$,  where $m_bv^\mu$ is
the {\it on-shell\/} part and $k^\mu$ is the {\it residual momentum.}  In this
decomposition, $k^\mu\sim\Lambda_{\rm QCD}$ represents the fluctuations in $p_b^\mu$ due
to the exchange of soft gluons with the rest of  the $B$ meson.  Only the on-shell part of
$p_b^\mu$ scales with $m_b$.  Also, mixing between the ``quark'' and ``anti-quark''
components of the Dirac spinor is suppressed by powers of $2m_b$, the mass gap between the
positive and negative energy parts of the wavefunction.  Hence  an effective heavy quark
field $h_v$ can be defined,
\begin{equation}
  h_v(x)={1+/\!\!\!v\over2}\,e^{im_bv\cdot x}\,b(x)\,,
\end{equation} where the Dirac matrix $(1+/\!\!\!v)/2$ projects out the ``quark'' part of 
the field. Furthermore, since
$i\partial^\mu h_v(x)=(p_b^\mu-m_b v^\mu)h_v(x)=k^\mu h_v(x)$, derivatives  acting on
$h_v$ scale as $\Lambda_{\rm QCD}$, rather than as $m_b$.

The next step is to express the QCD Lagrangian, ${\cal L}=\bar b(i/\!\!\!\!D-m_b)b$, in 
terms of the $m_b$-independent field $h_v$.  At lowest order in $1/m_b$, the result  is
\begin{equation}
\label{Chap2:eq:lhqet}
  {\cal L}_{\rm HQET}={\overline h}_v\, i v\cdot D h_v\,.
\end{equation} At leading order, ${\cal L}_{\rm HQET}$ respects the heavy spin and flavor
symmetries explicitly.  Both bottom and charm quarks can be treated as heavy by
introducing separate effective fields $h_v^{(b)}$ and $h_v^{(c)}$ and  duplicating ${\cal
L}_{\rm HQET}$.  The theory has a simple heavy quark propagator and  quark-gluon vertex
which are manifestly independent of $m_b$ and have no Dirac structure.

The effective theory is also expanded perturbatively in $\alpha_s(m_b)$.  In  particular,
the quark mass $m_b$ is shifted to $m_b^{\rm pole}$, the {\it pole  mass\/} at $n$ loops. 
The pole mass is a quantity which makes sense only at  finite order in perturbation
theory.  One must always be careful to be  consistent  in the convention by which one
chooses to define it.  

The mass of the $B$ meson may be expanded in powers of $m_b$,
\begin{equation}
  m_B=m_b+\Lbar+{\cal O}(1/m_b)\,,
\end{equation} where $\Lbar\sim\Lambda_{\rm QCD}$ is the energy contributed by the  light 
degrees of freedom.  Its precise definition depends on the convention by which one chooses
to define the heavy quark pole mass.  The parameter $\Lbar$ depends on the flavor,
excitation energy and total angular momentum of the  light  degrees of freedom. 

When one includes the leading $1/m_{b,c}$ corrections, the heavy spin and flavor
symmetries are broken by the subleading terms.  The leading
Lagrangian~(\ref{Chap2:eq:lhqet}) is modified by the addition of two terms,
\begin{equation}
  {\cal L}^{(1)}={1\over2m_b}\left(O_1+O_2\right)
  ={1\over2m_b}\left({\overline h}_v(iD)^2 h_v
  +{\overline h}_v\textstyle{1\over2}g\,G_{\mu\nu}\sigma^{\mu\nu}h_v\right)\,,
\end{equation} neglecting terms which vanish by the classical equations of motion.  Note
that the ``kinetic'' operator $O_1$ violates the heavy flavor symmetry, while the
``chromomagnetic'' operator $O_2$ violates both the spin and flavor symmetries.  When
radiative corrections are included, the operator 
$O_2$ is renormalized, and its coefficient develops a logarithmic dependence on
$m_b$.

The subleading operators $O_1$ and $O_2$ contribute to the mass  of the $B$ meson through
their expectation values,
\begin{eqnarray}
  \lambda_1&=&\langle B|\,{\overline h}_v(iD)^2 h_v\,|B\rangle/2m_B\,,\\
  \lambda_2&=&\langle B|\,{\overline h}_v\textstyle{1\over2}g\,G_{\mu\nu}
  \sigma^{\mu\nu}h_v\,|B\rangle/6m_B\,.
\end{eqnarray} The matrix elements $\lambda_1$ and $\lambda_2$ are often referred to by
the  alternate  names $\mu^2_\pi=-\lambda_1$ and $\mu_G^2=3\lambda_2$.  The parameter 
$\mu^2_\pi$  actually differs from $\lambda_1$ in that it is defined with an explicit 
infrared  subtraction. Because they are defined in the effective theory, the parameters 
$\lambda_1$ and $\lambda_2$ do not depend on $m_b$.  The expansion of
$m_B$ now takes the form
\begin{eqnarray}
\label{Chap2:eq:massexpand}
   m_B&=&m_b+\Lambda-{\lambda_1+3\lambda_2\over2m_b}+\dots\,,
   \nonumber\\
   m_{B^*}&=&m_b+\Lbar-{\lambda_1-\lambda_2\over2m_b}+\dots\,.
\end{eqnarray} Because ${\cal O}_2$ violates the heavy spin symmetry, it is the leading
contribution to the splitting between $B$ and $B^*$.  From the measured mass difference,
$\lambda_2\approx0.12\,{\rm GeV}^2$.  On the other hand, the parameters $\Lbar$ and
$\lambda_1$ must be measured indirectly.   Estimates from models yield the ranges
$200\,{\rm MeV}<\Lbar<700\,{\rm MeV}$ and $-0.5\,{\rm GeV}^2<\lambda_1<0$.  Measurement of
various features of inclusive semileptonic $B$ decays will provide experimental
information on
$\Lbar$ and $\lambda_1$ in the future.

\boldmath\subsection{Application of the HQE to $B$ Decays}\unboldmath
\label{Chap2:ssecBD}

There is a wide variety of applications of the HQE to $B$ decays.  Here  a few general
comments and two  illustrative examples are given.  In principle, the value of using an
effective theory such as the HQE is that there is a framework within which one can
estimate  the error in a calculation, due to uncomputed terms of a definite size.  Even
when the accuracy is not so good, it is under control in the sense that one can understand
the magnitude of the error to be expected.  In any application of  the HQE, then, two
sorts of questions must be addressed in addition to the computation itself:

1.\ What are the sizes of the leading uncomputed corrections in the expansion  in powers
of $\alpha_s$ and $1/m_b$ (or $1/m_c$, as appropriate)?  With what accuracy are the
parameters known which appear in the expansion?

2.\ What {\it other\/} assumptions or approximations have been made, beyond  those that go
into the HQE itself?

\subsubsection{Exclusive semileptonic $B$ decays}
\label{chap2:sssecBa}

The paradigmatic application of heavy quark symmetry is to semileptonic $B$  decay in the
limit $m_b,m_c\to\infty$.  This decay is mediated by the quark  transition $b\to
c\,\ell\bar\nu$.  Suppose the weak decay occurs at time $t=0$. What  happens to the light
degrees of freedom?  Since the $b$ quark does not recoil, for 
$t<0$ they  see simply the color field of a point source moving with velocity $v$.  At
$t=0$, this point source changes  (almost) instantaneously to a new velocity $v'$; the
color neutral leptons do not  interact with the light hadronic degrees of freedom as they
fly off.  The light quarks and gluons then must reassemble themselves about the recoiling
color source.  There is some chance that this nonperturbative process will lead the light
degrees of freedom to reassemble themselves back into a $D$ meson.  The amplitude for this
to happen is a function $\xi(w)$ of the product $w=v\cdot v'$ of the initial and final
velocities of the heavy color sources.

Clearly, the kinematic point $v=v'$, or $w=1$, is a special one.  In this  corner of phase
space, where the leptons are emitted back to back, there is no  recoil of the source of
color field at $t=0$.  As far as the light degrees of freedom are concerned, {\it nothing
happens!\/}  Hence the amplitude for them to remain in the ground state is exactly unity. 
This is reflected in a nonperturbative normalization of $\xi(w)$ at zero recoil,
\begin{equation}
\label{Chap2:eq:norm}
  \xi(1)=1\,.
\end{equation} This normalization condition is of great phenomenological use.  There are
important corrections to this result for finite heavy quark masses $m_b$ and, especially,
$m_c$.

The weak decay $b\to c$ is mediated by a left-handed current 
$\bar c\gamma^\mu(1-\gamma^5)b$, which can also change the orientation of the  spin  $S_Q$
of the heavy quark during the decay.  Since the only difference between  a $D$  and a
$D^*$ is the orientation of $S_c$, heavy quark symmetry implies  relations  between the
hadronic matrix elements which describe the semileptonic decays 
$B\to D\ell\bar\nu$ and $B\to D^*\ell\bar\nu$. These matrix elements are parameterized by
six form factors, which are  independent  nonperturbative functions of $w$.  In the heavy
quark limit, they are all  proportional  to $\xi(w)$, a powerful constraint on the
structure of semileptonic decays.

Now consider more closely the structure of the theoretical expansion for  the decay  $B\to
D^*\ell\bar\nu$, which may be used to measure the CKM matrix element
$|V_{cb}|$.  Near the zero-recoil point, the transition is dominated by a single form
factor, $h_{A_1}(w)$, with the normalization condition $h_{A_1}(1)=1$ in the heavy quark
limit.  For general $m_B$ and 
$m_{D^*}$,  the differential decay rate may be written 
\begin{equation}
   {{\rm d}\Gamma\over{\rm d}w}=G_F^2\,|V_{cb}|^2\,K(m_B,m_{D^*},w)\,F^2(w)\,,
\end{equation} where $K(m_B,m_D,w)$ is a known kinematic function and $F(w)$ has an
expansion at $w=1$ of the form
\begin{equation}
  F(1)=\eta_A(\alpha_s)\left[1+{0\over m_c}+
  {0\over m_b}+{\cal O}\left(1/m_b^2\right)\right]\,.
\end{equation} The perturbative function $\eta_A(\alpha_s)$ has been computed to two
loops,  with the result $\eta_A=0.960$.  The leading HQE corrections arise at order 
$1/m^2_{b,c}$  rather than at order $1/m_{b,c}$, and have been estimated  to be
approximately  5\%.  More detailed analysis of this decay exist in the literature.  The
point here is to note how the double  expansion in powers of $\alpha_s$ and $1/m_{b,c}$
appears in a physical quantity.  This analysis is also typical because it applies only to
a very particular case of enhanced symmetry, namely the decay rate as
$w\to1$.  The extrapolation  of the data to this limit requires both experimental
ingenuity and more theoretical input beyond the HQE.

\subsubsection{Duality and inclusive semileptonic decays}
\label{Chap2:subs:du}

As a second example, consider the {\it inclusive\/} decay $B\to X_c$, where 
$X_c$ is any final state containing a charm quark.  The analysis of inclusive decays, 
although it  relies on a similar expansion, is different from the treatment of exclusive 
decays.   In this case, it is useful to observe that the energy released into the final
state by the decay of the heavy
$b$ quark is large compared to the QCD scale.  Hence the final hadronic state need not be
dominated by a few sharp resonances. If resonances are indeed unimportant, then there is a
factorization between the short distance part of  the decay (the disappearance of the $b$
quark) and the long distance part (the eventual hadronization of the decay products). 
This factorization implies that for sufficiently inclusive quantities it is enough to
consider the short distance part of the process, with the subsequent hadronization taking
place  with unit probability.  Note that what is important here is that the
$b$ quark is heavy, with no restriction placed on the charm mass.  In fact, a smaller
charm quark mass is better, because it increases the average kinetic energy of the decay
products.

This factorization, known as {\it local parton-hadron duality}, is an example  of a 
crucial assumption which lies outside of the HQE itself.  What is its status?   Clearly,
local duality must hold as $m_b\to\infty$ with all other masses held fixed.  In this
limit, wavelengths associated with the $b$ quark decay are arbitrarily short and cannot
interfere coherently with the hadronization process.  On the other hand, it is not known
how to estimate the size of corrections to local duality for $m_b$ large but finite. 
There is no analog of the heavy quark expansion appropriate to this question, and no way to
estimate systematically deviations from the limit $m_b\to\infty$.  Although we will
incorporate an expansion in powers $1/m_b$ in the calculation of inclusive quantities, the
behavior of this expansion does not address directly the issue of violations of duality. 
The duality hypothesis, while entirely reasonable for inclusive $B$ decays, is not
independently verifiable except by the direct confrontation of theoretical calculations
with the data.

For semileptonic $B$ decays, $B\to X_c\,\ell\,\bar\nu$, the situation has  additional 
interesting features.  On the one hand, in the region of phase space where the leptons
carry away most of the available energy, the final hadronic state is likely to be
dominated by resonances and local duality is likely to fail.  (In some $B$ decays, local
duality can be shown to hold even in the resonance region; however, this requires a more
subtle and less intuitive argument than the one on which this dicussion is based.)  On the
other hand, if one integrates over the lepton phase space to compute an  inclusive
quantity such as the total semileptonic width, then one needs not local  duality but 
rather the weaker notion of {\it global parton-hadron duality.}  (The use here and
elsewhere in this book of this term, while it reflects current practice, is ahistorical. 
This notion originally was known simply as {\em duality}, while {\em global duality\/} was
first introduced to describe the technical assumption that one can neglect distant cuts in
computing the semileptonic $B$ decay rate, which is an important and distinct issue.  Both
terminologies  remain in use in the literature.) In essence,  the argument is as follows. 
Let $q$ be the momentum carried away by the leptons.  The semileptonic width is an
integral of a differential width, written schematically as 
${\rm d}\Gamma/{\rm d}q$,  which must be calculated under the hypothesis of local
duality.  For certain ranges of
$q$ ($q^2$ near its kinematic maximum), local duality clearly fails.  However, 
${\rm d}\Gamma/{\rm d}q$ has a known analytic structure as a function of $q$,  with cuts 
and poles, corresponding to thresholds and resonances,  which are confined to the real 
axis.   If the integration contour in $q$ is deformed away from the resonances, {\it into
the complex plane,} then it may be possible to compute  the integral without knowing the
integrand everywhere along the original (real) contour of integration.  From one point of
view, complex $q$ forces the final state away from the mass shell, where long distance
effects can become important.  From another, the integral over $q$ imposes an average over
the invariant mass $s_H$ of the hadrons in the final state, which smears out the effect of
resonances when they do contribute.  This property, that quantities averaged over $s_H$
may be computable even when differential ones are not, is global duality.  The most
important feature is the {\it smearing\/} of the  perturbative  calculation over the
resonance region.  Note that global duality does not  apply to  purely hadronic $B$
decays, for which $s_H=m_B^2$ is fixed.

Once the issue of duality has been addressed, the actual expansions obtained  for 
inclusive decays are very similar to those for exclusive decays.  For example,  the total
charmless semileptonic $B$ decay width takes the form
\begin{equation}
  \Gamma(B\to X_u\,\ell\bar\nu)
  ={G_F^2|V_{ub}|^2\over192\pi^3}\,m_b^5\left[1-2.41{\alpha_s\over\pi}+
  {\lambda_1-9\lambda_2\over2m_b^2}+\dots\right]\,.
\end{equation} The leading corrections to this expression are of order $\alpha_s^2$, 
$\alpha_s/m_b$ and $1/m_b^3$.  Note that the $1/m_b^2$ corrections are far more tractable 
than in the exclusive decay: first, because $1/m_c^2$ does not appear, and second, because
they may all be written in terms of the two parameters $\lambda_1$ and
$\lambda_2$, one of which is already known.  Finally, note the strong  dependence on the
mass $m_b$, which is equivalent via the mass expansion  (\ref{Chap2:eq:massexpand}) to  a
dependence on $\Lbar$.  This is a significant source of uncertainty in the  expression for
the rate.  There have been extensive recent theoretical efforts to reduce this
uncertainty.  The  technically more complicated case of $B\to X_c\,\ell\bar\nu$
is discussed in the literature.

\subsection{Limitations of the HQE}
\label{Chap2:ssecBE}

While the heavy quark expansion and the HQET are powerful tools in the analysis of many
aspects of $B$ spectroscopy and decay, there are important issues into which they provide
little direct insight.  What the HQE provides is a framework within which the dependence
of quantities on the large mass $m_b$  may be extracted systematically.  However, once
this has been accomplished, the task usually remains of analyzing those parts of the
process which are  characterized by long distances, small momenta, and nonperturbative
dynamics.  For a few quantities, such as the exclusive and inclusive decay rates discussed
in this section, the calculation can be organized so that such effects appear only at
subleading order, with the leading order terms controlled by heavy quark symmetry.  But
this is not the typical situation in $B$ phenomenology; one is usually required to analyze
quantities and processes for which the nonperturbative nature of QCD is a dominant effect.

By necessity, such analyses involve a wide variety of methods, techniques, approximations
and ansatzes.  Some of the most  important approaches are briefly discussed in the rest of
this introduction.  But even  where the heavy quark limit is not itself predictive, it
still has an important  role to play.  Any model or effective theory which purports to
describe $B$ mesons  must obey the heavy quark  limit.  By the same token, it is often
possible to enhance or extend a model by building heavy quark symmetry in explicitly.  The
information provided by the heavy quark limit will prove to be very useful in this broader
context.

\section{Light Flavor Symmetry}
\label{Chap2:secCC}

\subsection{Chiral Lagrangians}
\label{Chap2:ssecCA}

Complementary to the heavy quark limit, new symmetries of QCD also arise in the limit of
vanishing light quark masses. As $m_u,\,m_d,\,m_s\to0$, the quarks  of left and right
helicity decouple from each other.  In this limit, the invariance  of the Lagrangian
separately under rotations among $(u_L,d_L,s_L)$ and  
$(u_R,d_R,s_R)$ gives rise to an $SU(3)_L\times SU(3)_R$ chiral flavor symmetry.  In the
QCD vacuum, this symmetry is dynamically broken to the diagonal subgroup 
$SU(3)_V$ by the quark condensate $\langle \bar q _iq_j\rangle\ne0$, . As a consequence,
there are eight Goldstone bosons in the light spectrum, which we identify with the
physical $\pi$, $K$, and $\eta$. Since the actual $u$, $d$,  and $s$ quark masses are
small but nonzero, the $\pi$, $K$, and $\eta$ are light but not exactly massless. 

The spontaneous breaking of chiral symmetry is characterized by a scale
$\Lambda_\chi\sim1\,$GeV, which is related to the value of the quark  condensate. For
light masses $m_\pi,m_K,m_\eta\ll\Lambda_\chi$ and small momenta
$p\ll\Lambda_\chi$, QCD exhibits a separation of energy scales which can be  used as the
basis for an effective field  theory. This {\it chiral Lagrangian}  describes low energy 
interactions in a systematic expansion in powers of
$p/\Lambda_\chi$ and $m_q/\Lambda_\chi$.  Since the fundamental degrees of freedom, the
eight ``pions'' $\pi$, $K$ and $\eta$, are the Goldstone bosons  associated with the
spontaneous symmetry breaking 
$SU(3)_L\times SU(3)_R\to SU(3)_V$,  they transform in a complicated nonlinear way under
the full symmetry group.   It is convenient to assemble them into a matrix $\Pi_{ab}$,
which is in turn  exponentiated,
\begin{equation}
  \label{Chap2:sigmafield}
  \Sigma_{ab} = \left[\exp{\left( 2 i \Pi/f_\pi \right)}\right]_{ab}\,,
\end{equation} where $f_\pi\approx130\,$MeV is the pion decay constant and $a,b$ are
flavor indices which take values $u,d$ or $s$.  The unusual looking field
$\Sigma$ has the property that it transforms simply under 
$SU(3)_L\times SU(3)_R$.

The chiral Lagrangian is the most general function of $\Sigma$ consistent  with the
symmetries, constructed order by order in powers of $1/\Lambda_\chi$.   At lowest order,
the Lagrangian is completely fixed,
\begin{equation}
\label{Chap2:CPTlagrangian}
   {\cal L} = { f_\pi^2 \over 4} \,
   \partial^\mu \Sigma_{ab}^\dagger \partial_\mu \Sigma_{ba} 
   + \ldots \ ,
\end{equation} where the flavor indices are summed over and  the ellipsis indicates
operators suppressed by $\Lambda_\chi^n$.   The exponential form of the $\Sigma$ field
allows this simple Lagrangian to describe  interactions between arbitrarily large numbers
of pions. Indeed, one of the useful features of Chiral Perturbation Theory (ChPT) is its
ability to  relate scattering amplitudes involving different numbers of external particles.

All hadrons other than the pions, such as vector mesons or baryons,  have masses of order
$\Lambda_\chi$.  Hence for external momenta 
$p\ll\Lambda_\chi$ they can  only appear as virtual states.  Their effect on the effective
theory is   reproduced by higher dimension operators involving $\Sigma$, such as
\begin{equation}
  {f_\pi^2\over\Lambda_\chi^2}\,{\rm Tr}\,\left[\partial^\mu\Sigma^\dagger
  \partial^\nu\Sigma^\dagger\partial_\mu\Sigma\partial_\nu\Sigma\right]\,,
  \nonumber
\end{equation} where the trace is over flavor indices.  Because one cannot solve QCD, the
couplings of these operators are unknown constants which must be determined
phenomenologically. In practice, the Lagrangian (\ref{Chap2:CPTlagrangian}) has been
generalized to include operators containing up to four derivatives or  one power of the
light quark masses, as well as effects from electromagnetic and flavor changing currents.
By now, most of the couplings have been extracted from  experiment. Typical predictions,
such as $\pi$--$K$ radiative reactions or
$\pi$--$\pi$ scattering, are accurate at the $10$--$30\%$  level, although in  some cases,
such as the extraction of  $|V_{us}|$ from $K \to \pi \ell \bar\nu$,  the uncertainties
are much smaller.  It is important to keep in mind that these  predictions   are valid
only so long as external momenta are small compared to 
$\Lambda_\chi\sim1\  {\rm GeV}$.

\subsection{Heavy Hadron Chiral Perturbation Theory}
\label{Chap2:ssecCB}

Although heavy hadrons have masses much larger than $\Lambda_\chi$, it is still possible
to incorporate them into ChPT.  This is because it is only the light degrees of freedom in
the hadron, whose mass does not scale with the heavy  quark, which interact with external
pions.  This extension of the effective theory,  known as Heavy Hadron Chiral Perturbation
Theory (HHChPT), incorporates the heavy  quark spin-flavor  symmetry in an expansion in
derivatives, light quark masses, and inverse heavy quark masses.  It describes soft pions
interacting with a static heavy hadron.

A simple example of where such a formalism is useful is the semileptonic decay 
$B \to \pi\ell \bar\nu$. Over most of the Dalitz plot, the pion is much too energetic for
chiral symmetry to apply. However, in the region where the pion is soft, the form factor
$f_+(q^2)$ which determines the differential  rate can be determined reliably. For a
sufficiently soft pion, the dominant  contribution to 
$B \to \pi\ell \bar\nu$ comes from the process where $B\to B^*\pi$, with the virtual $B^*$
then decaying leptonically.  The strength of the $B\to B^*\pi$  transition  is
proportional to a universal coupling constant $g$, which may be determined  from the  rate
for the decay $D^*\to D\pi$.  The amplitude for $B \to \pi\ell \bar\nu$  at lowest  order
in HHChPT is then
\begin{equation}
  f_+(q^2) = {g M_B^2 f_B/f_\pi \over M_{B^*}^2-q^2}\ ,
\end{equation} which is simply a statement of nearest pole dominance, which holds
rigorously in the combined heavy quark and chiral limit. Physically, pole dominance holds
because in this limit the mass splitting between the $B$ and 
$B^*$ vanishes, whereas the energy gap to the nearest excited resonance remains  finite.
Thus, for arbitrarily soft pions,  the $B^*$ is the only resonance which can  affect the
form factor. Note that the heavy and light flavor symmetries relate 
$B_s, D$ and $D_s$ states to the $B$, so there are analogous form factors in 
$B_s \to K \ell \bar\nu, D_s \to K \ell \bar\nu$, and $D \to \pi \ell \bar\nu$.

As is typical in chiral calculations, the amplitude relations hold only where  the  pions
are soft.  There will be corrections to these relations at higher order,  when loop graphs
and explicit symmetry breaking terms are included.  Most  calculations within HHChPT are
done at  leading order, or include only some of  the numerically important corrections. 
Since the number of unknown  coefficients  tends to proliferate at higher order, such
results are usually presented as  estimates of the size of symmetry breaking effects. 
Chiral Lagrangians are  particularly useful for exploring the light flavor dependence of
quantities  arising from pion loops and other infrared physics.

\subsection{Factorization, Color Flow, and Vacuum Saturation}
\label{Chap2:ssecCC}

The problem with chiral calculations is that they only apply when the  external pions  are
soft, and for most processes of phenomenological interest, nothing  constrains  this to be
the case.  For example, it is not very useful to apply such  techniques  to exclusive
nonleptonic decays such as $B\to D\pi$, since the $\pi$ has  momentum  $p=2.3\,{\rm
GeV}>\Lambda_\chi$ .  If one attempts to use the chiral Lagrangian  here, one  finds that
the effects of higher dimension operators, which scale as 
$(p/\Lambda_\chi)^2 $, are unsuppressed, and the theory loses its  predictive power.   The
hadrons in the final state continue to interact long after the weak  decay, and  there is
no clean separation of scales.  The situation is even more  complicated for  multiple pion
production ($B\to D\pi\pi,\cdots$), which is governed over  most of the  phase space not
by low energy theorems but by the nonperturbative dynamics  of QCD  fragmentation.  In the
absence of a solution to QCD, exclusive nonleptonic  decays  remain one of the most
intractable problems in $B$ physics.  All that one has is a  variety of models, based on
ideas such as light cone wavefunctions or  fragmenting  strings, which describe the data
with varying degrees of success. 

In the absence of any theory based on first principles, phenomenological  approaches are
often used instead.  The most popular of these is the  hypothesis of {\it factorization,}
which applies to certain two body  nonleptonic decays.  A simple example is $B\to D\pi$,
which is mediated by  the quark transition $b\to c\bar u d$.  Immediately after the weak
decay,  the quarks typically find themselves with a large momentum and in the  middle of a
medium of gluons and light quark-antiquark pairs, with which  they subsequently interact
strongly.  However, if the $\bar u d$ pair has a  small invariant mass, $m(\bar u d)\approx
m_\pi$, then these two quarks will  remain close together as they move through the colored
medium.  If, in  addition, they are initially in a color singlet state, then they will
interact  with the medium not individually but as a single color dipole.  Since the 
distance between the $\bar u$ and the $d$ grows slowly, it is possible that  the pair will
have left the colored environment completely before its dipole  moment is large enough for
its interactions to be significant.  In this case,  the pair will hadronize as a single
$\pi$.  Such a phenomenon is known as  ``color transparency''.

If, by contrast, the $\bar u d$ pair has a large invariant mass, then the  quarks will
interact strongly with the medium.  In  this case,  their reassembly into a single $\pi$
is extremely unlikely.  As a result,  it is  reasonable to hypothesize that the decay
$B\to D\pi$ is dominated by the  former  scenario, and that the matrix element actually
factorizes,
\begin{equation}
\label{Chap2:eq:factorization}
  \langle D\pi|\,\bar c\gamma^\mu(1-\gamma^5)b\,
  \bar d\gamma_\mu(1-\gamma^5)u\,|\bar b\rangle
  =\langle D|\,\bar c\gamma^\mu(1-\gamma^5)b\,|\bar b\rangle\times
  \langle\pi|\,\bar d\gamma_\mu(1-\gamma^5)u\,|0\rangle\,.
\end{equation} The result is something much simpler:  
$\langle\pi|\,\bar d\gamma_\mu(1-\gamma^5)u\,|0\rangle$ is related to 
$f_\pi$, and 
$\langle D|\,\bar c\gamma^\mu(1-\gamma^5)b\,|\bar b\rangle$ may be extracted from 
semileptonic $B$ decays.  With this ansatz, it is possible to obtain relations  among
various two body decays which can then be tested experimentally.   A proper  analysis is
fairly complicated, because it is necessary to take into account  short distance
perturbative corrections and other formally subleading  effects.   In particular, when the
leading QCD radiative corrections are included, the matrix element (2.31) develops a
dependence on the renormalization scale $\mu$ which cannot be compensated within the
factorization ansatz. Thus even the question of whether a matrix element factorizes has no
scale invariant meaning.

There is a heuristic distinction which is often made in the discussion of  nonleptonic $B$
decays, between contributions to decays which are  ``color allowed'' and those which are
``color suppressed''.  In the spirit  of factorization, it is often convenient to use
Fierz identities to rewrite the  effective Hamiltonian as a sum of products of quark
bilinears which  could interpolate certain exclusive final states.  For example, if one 
were interested in the semi-inclusive process $B\to X_s\psi$, it would  be useful to
re-express the combination
\begin{equation}
  C_1\,\bar s_i\gamma^\mu(1-\gamma^5)c^j\,
  \bar c_j\gamma_\mu(1-\gamma^5)b^i+
  C_2\,\bar s_i\gamma^\mu(1-\gamma^5)c^i\,
  \bar c_j\gamma_\mu(1-\gamma^5)b^j\,,
\end{equation} where $i$ and $j$ are color indices, as
\begin{equation}
  (C_1+\textstyle{1\over3}C_2)\,\bar c\gamma^\mu(1-\gamma^5)c\,
  \bar s\gamma_\mu(1-\gamma^5)b+
  2C_2\,\bar c\, T^a\gamma^\mu(1-\gamma^5)c\,
  \bar s\, T^a\gamma_\mu(1-\gamma^5)b\,.
\end{equation} Then the first term can be factorized in the sense of
Eq.~(\ref{Chap2:eq:factorization}),  while the second cannot.  If the coefficient
$C_1+{1\over3}C_2$ of the  factorizable term is large, that is, if $C_1+{1\over3}C_2\gg
2C_2$, then  the amplitude is said to be ``color allowed''; if the reverse is true, then
it is  said to be ``color suppressed''. It is often supposed that amplitudes which  have
the wrong color structure to factorize are intrinsically small.  Of course,  soft gluons
can always be exchanged to rearrange the color structure, so this  distinction does not
survive radiative corrections.  However, the neglect of  nonfactorizable amplitudes is a
common phenomenological starting point for  analyses of nonleptonic $B$ decays, where it
is often useful to have some  guess as to which four-quark operators are the most
important for  mediating a given transition.

Another common ansatz, which is similar in spirit to factorization, is  {\it vacuum
saturation.}  The computation of $B^0$--$\bar b^0$  mixing requires the hadronic matrix
element 
$\langle \bar b^0|\,\bar b\gamma^\mu\gamma^5d\,\bar b\gamma_\mu
\gamma^5d\,|B^0\rangle,$ where the four quark operator has been induced  by an interaction
(such as a box diagram) at very short distances.  In  vacuum saturation, one inserts a
complete set of states between the two  currents, and then assumes that the sum is
dominated by the vacuum.  This  ansatz is neither stable under radiative corrections, nor
really well  defined, since $\bar b\gamma^\mu\gamma^5d\,\bar b\gamma_\mu\gamma^5d$ is  an
indivisible local operator.  The result is of the form
\begin{equation}
  \langle \bar b^0|\,\bar b\gamma^\mu\gamma^5d\,
  \bar b\gamma_\mu\gamma^5d\,|B^0\rangle=A f_B^2 m_B^2 B_B\,,
\end{equation} where $A$ is a known constant and $B_B$ absorbs the error induced by
keeping  only the vacuum intermediate state.  Deviations of $B_B$ from unity 
parameterize  corrections to the ansatz.  Vacuum saturation becomes exact in the formal 
limit  $N_c\to\infty$, where $N_c$ is the number of colors, since then the mesons are 
noninteracting.  This limit is often cited as a justification of the ansatz.   As it turns
out, calculations in lattice QCD do seem to prefer a value for 
$B_B$  which is close to unity.  One may use HHChPT to estimate the uncertainty in  the 
light flavor dependence of the ratio $B_{B_s}/B_{B_d}$.

\section{Lattice Gauge Theory}
\label{Chap2:secDD}

An important alternative to the analytic analyses presented so far is the  attempt  to
solve QCD directly via a numerical simulation.  As for any quantum field  theory,  QCD may
be defined by a partition function,
\begin{equation}
  Z=\int[{\rm d}A_\mu][{\rm d}\overline\psi_i][{\rm d}\psi_i]\,
  e^{iS(A_\mu,\overline\psi_i,\psi_i)}\,,
\end{equation} where the {\it functional integral\/} is over all configurations with given 
gauge  potential $A_\mu$ and quarks $\psi_i$.  The action,
\begin{equation}
   S(A_\mu,\overline\psi_i,\psi_i)=\int{\rm d}^4x\left[
   -\textstyle{1\over4}\,G^{\mu\nu}G_{\mu\nu}
   +\overline\psi_i(i/\!\!\!\partial-g\,/\!\!\!\!A-m)\psi_i+\dots\right]\,,
\end{equation} is supplemented by sources for the quarks and gluons, and by gauge fixing 
terms.   The functional $Z$ and its derivatives are enough to determine all of the 
correlation functions of the theory.  The program of  {\it lattice gauge theory\/}  is to
perform the integral in the action by  discretizing space-time on a grid  of  spacing $a$,
and then to compute $Z$ by summing over a finite but  representative  set of
configurations of $A^\mu$ and $\psi_i$.  In principle, given enough  configurations and a
fine enough grid, such an analysis provides an  arbitrarily  accurate solution to QCD.

However, there are a number of important practical difficulties with this  program,  which
effectively restrict its accuracy and rigor, and the uses to which it  may be  put.  The
first is that any realistic analysis requires an enormous amount of  computer power. 
While such resources continue to improve at a remarkable  pace,  it will be long in the
future  before it will be possible to analyze  processes in which a wide range of momentum
scales is important.  Effectively,  this limits the use of the lattice for the study of
exclusive nonleptonic $B$  decays or $\pi$--$\pi$ scattering.  For the time being, it is
the static,  rather  than the dynamical, properties of QCD which are most amenable to a
lattice  treatment.

Another practical limitation of lattice QCD is that it is extremely  expensive  to 
include quark loops in the computation.  It is possible to save a huge  factor in 
computing time by working in the {\it quenched \/} or {\it valence  approximation\/},  in
which quark loops are neglected entirely.  Quenching is really more an  ansatz  than an
approximation, in the sense that it is difficult to estimate reliably  the  error which it
induces.  It can be argued that in certain contexts, such as  heavy  quark-antiquark bound
states, the primary effect of quenching is to  renormalize the  effective coupling of the
gluons, which can be compensated by adjusting the  coupling $g$ at the lattice scale.  But
in most cases, quenching is just a  necessary simplification of the calculation, with a
largely unknown effect on  the results.  With the emergence of a new generation of
computers capable of 
$\sim10^{12}$ flops, some unquenched calculations will become feasible.  Then  it will
begin to be possible to study the effects of quenching in more  quantitative detail.

Other practical difficulties in lattice QCD are more tractable.  Because of  the  nature
of the propagator, massless quarks induce singularities in lattice  calculations, so one
must work with light quarks of mass $m\sim 100\,$MeV or  larger and then extrapolate to
physical $m_u\sim5\,$MeV and $m_d\sim10\,$MeV.   The nature of this extrapolation is
strongly affected by quenching.  It is  also  necessary to work at nonzero lattice spacing
$a$, and finite overall lattice  size $L$, extrapolating to $a\to0$ and $L\to\infty$ at
the end.  These  extrapolations are believed to be reasonably under control in most 
calculations.   Finally, it turns out to be extremely difficult to incorporate chiral 
quarks in  lattice computations, although this is not an important problem for a vector 
theory such as QCD.

Even given these limitations, the progress in lattice QCD in the past  ten years  has been
phenomenal.  This is due both to advances in computing technology,  and  perhaps more
important, to the development of new theoretical methods  particular  to the lattice.  New
techniques which are relevant to the study of  heavy quarks  include the {\it static
approximation, nonrelativistic QCD,} and {\it improved  actions.}  The first of these is
the analogue of HQET for heavy quarks on the  lattice, which actually predates (and
inspired) the development of HQET in the  continuum.  Static techniques are necessary
because the Compton wavelength of  the $b$ quark scales as $1/m_b$ and is much smaller
than any lattice spacing 
$a$  in use, so fully dynamical $b$ quarks are extremely difficult to simulate.  The 
static limit has proven very useful for the computation of heavy hadron  spectra  and
decay constants.  Nonrelativistic QCD, a somewhat different expansion in  powers of $1/m$,
is relevant to the study of heavy quark-antiquark bound  states.   Such analyses have
become so accurate that lattice determinations of  quarkonium  splittings, when compared
with data, provide a measurement of $\alpha_s$ which  may be competitive with precision
measurements at the $Z$ pole.  Finally,  it has  been understood how to ``improve'' the
action 
$S(A_\mu,\psi_i,\overline\psi_i)$  by including discretization effects order by order in
$a$, thereby allowing  the  same accuracy to be obtained with larger lattice spacing. 
Since for a lattice  of a given size in physical units, the number of points scales as
$1/a^4$, the  result can be a significant saving in computer resources.

In summary, the lattice will continue to be an important tool for $B$ physics,  but it is
not a universal approach for the numerous important quantities which  cannot as yet be
treated analytically.  Lattice QCD has been very successful  for  certain quantities, such
as the bag constant $B_B$ and the splittings in the 
$\Upsilon$ system.  For others, such as bottom meson and baryon spectroscopy,  decay
constants, and semileptonic form factors, the situation continues to  improve.  At the
same time, there are quantities, such as exclusive nonleptonic decays  or  fragmentation
functions, where lattices of impractical size and granularity  would  be required to
obtain useful predictions.  For such processes, there is  no rigorous theoretical
calculation based on first principles.  

\section{QCD Sum Rules}
\label{Chap2:secEE}

Another theoretical approach which is based, in principle, on QCD is that of {\it QCD sum 
rules.}  The idea is to exploit parton-hadron duality as fully as possible, by  studying
inclusive quantities with kinematic or other restrictions which  require  them to be
dominated by a single exclusive intermediate state.  In this way  one  can learn something
about nonperturbative physics, which controls the detailed  properties of bound states,
within a calculation based on a perturbative  expansion.

The construction of QCD sum rules has a number of technical subtleties.   This  section 
only outlines the ingredients and the general structure.   The method involves the study
of  correlation functions in QCD, as a function  of external momenta.  The  correlation 
functions of a quantum field theory contain all the  information about the  theory,  and
hence in principle this method has access to every observable of QCD.  In  practice, it is
only feasible to study two-point and some three-point  functions,  so QCD sum rules are
useful primarily for computing spectra, decay constants  and  form factors.

The main features of the method will be illustrated here with a simple example from 
$B$  physics.  One attractive feature of sum rules is that they can be formulated  within 
HQET, thereby incorporating heavy quark symmetry automatically.  In HQET, the  current
${\overline h}_v\gamma^5 q$ can create a $B$ meson, or other excited states 
$B_n$ with the same quantum numbers.  Since the $b$ quark is static, the  appropriate
measure of the mass of a state is the excitation energy 
$\nu_n=(m_{B_n}^2-m_b^2)/2m_b$, which is independent of $m_b$ as 
$m_b\to\infty$.   The object of study is the two point function
\begin{equation}
\label{Chap2:eq:correlator}
   \Pi(\omega)=i\int{\rm d}^4x\,e^{ik\cdot x}\,
   \langle0|\,T\{\bar q \gamma^5 h_v(x),{\overline h}_v\gamma^5 q(0)\}\,|0\rangle\,,
\end{equation} where $\omega=v\cdot k$ is the energy injected into the correlator.  For 
general $\omega$, the correlator $\Pi(\omega)$ receives contributions from  all
intermediate states $B_n$.  Hence  $\Pi(\omega)$ may be written in the form
\begin{equation}
\label{Chap2:eq:statesum}
  \Pi(\omega)=\sum_n{F_n^2\over\nu_n-\omega-i\epsilon}\,,
\end{equation} where $F_n$ is the coupling of the current to the excited state $B_n$.  The 
correlator~(\ref{Chap2:eq:correlator}) and sum over  states~(\ref{Chap2:eq:statesum}) are 
often referred to, respectively, as the ``theoretical'' and  ``phenomenological''  sides
of the sum rule.

The goal is now two-fold:  first, to compute the  correlator~(\ref{Chap2:eq:correlator}) 
in QCD, and second, to isolate the contribution of the ground state $B$ to the 
sum~(\ref{Chap2:eq:statesum}), so that its properties can be extracted.   These two  goals
conflict, as they require different limits of $\omega$.  A perturbative  calculation of
the correlator is appropriate for $\omega$ far from resonances, 
$\omega\gg\Lambda_{\rm QCD}$, or even better, in the unphysical region 
$\omega\to-\infty$.  On the other hand, the ground state will only dominate  the sum for
$\omega$ small and near the $B$ resonance.  The compromise is to  work in a regime of
intermediate $\omega$, where it is hoped that, with some  technical improvements, both the
correlator and the sum over states can be  treated accurately.  These improvements are the
source of most of the  complications in the method.

The first step is to rewrite $\Pi(\omega)$ as a dispersion integral over its  imaginary
part, which receives contributions from real intermediate states,
\begin{equation}
\label{Chap2:eq:dispersion}
  \Pi(\omega)=\int_0^\infty{\rm d}\nu{\rho(\nu)\over
  \nu-\omega-i\epsilon}\,,
\end{equation} where $\rho(\nu)\propto{\rm Im}\Pi(\nu)$.  For simplicity, local 
subtractions, which may be required to make this expression well behaved,  have been
omitted here.  Note the similarity between the theoretical expression 
(\ref{Chap2:eq:dispersion})  and the phenomenological sum over states 
(\ref{Chap2:eq:statesum}).   While it is  certainly not true that 
$\rho(\nu_n)=F_n^2$ at each point, global duality  allows  the two expressions for
$\Pi(\omega)$ to coincide once both sides have been  integrated.  For $\omega$ large
enough, it suffices to compute the density 
$\rho(\nu)$ as a power series in $\alpha_s$.  But for intermediate $\omega$,  it is
necessary also to include corrections to $\Pi(\omega)$ of order 
$1/\omega^n$.  These corrections appear in the form of {\it condensates\/}, new 
nonperturbative  quantities characteristic of QCD.  It is usually enough to include the 
condensates  of dimension $\le5$, whose values have been extracted at the $\sim30\%$ 
level from  other processes:
\begin{eqnarray}
  &&\langle\bar q  q\rangle\approx-(0.23\,{\rm GeV})^3\,,\nonumber\\
  &&\langle\alpha_sG^{\mu\nu}G_{\mu\nu}\rangle\approx(0.45\,{\rm GeV})^4\,,
  \nonumber\\
  &&\langle g\bar q \sigma^{\mu\nu}q\,G_{\mu\nu}\rangle
  \approx-(0.40\,{\rm GeV})^5\,.
\end{eqnarray} The condensates are universal quantities which, it is hoped, capture the 
leading  nonperturbative effects of the QCD vacuum and allow the correlator to be 
computed  accurately even for $\omega$ not asymptotically large.

The next step is to focus on the ground state $B$.  The excited states $B_n$  are all 
quite broad and unlikely to induce rapid variations in $\Pi(\omega)$, and it  is  assumed
that above some scale $\omega_0$, the integral over the excited states  can accurately be
described by parton-hadron duality.  Actually, it is hoped  that the scale $\omega_0$ may
be chosen as a {\it threshold,} in the sense  that  the entire contribution of the excited
states (and none of the ground state)  may  be modeled by the perturbative dispersion
integral for $\nu>\omega_0$.  The  contribution of the excited states is then subtracted
from both expressions,  leaving an upper cutoff $\omega_0$ on the dispersion 
integral~(\ref{Chap2:eq:dispersion}),  and only the ground state $n=0$ in the  sum over
intermediate  states~(\ref{Chap2:eq:statesum}).

The object of the analysis is now to equate the theoretical and  phenomenological  sides
of the sum rule and  attempt to fit the coupling $F$ and the  energy $\nu$  of the ground
state $B$.  To do so, one must fix values for the threshold 
$\omega_0$  and the energy $\omega$, neither of which is given {\it a priori.}  (In Borel 
sum rules, $\omega$ is exchanged for a ``Borel parameter'' $T$.)  While there  exists a
prescription for choosing these parameters, it is not based directly  on  QCD, but rather
derives from the requirement that the sum rule be  self-consistent,  that is, dominated
neither by the condensates nor by the excited states.   In fact,  therein lies a
fundamental source of uncertainty in the practical application of  QCD sum rules.  While
it is certainly encouraging that $\omega_0$  and $\omega$  usually may be chosen to make
the sum rule consistent and well behaved,  there is  no way to test whether the {\it
consistent\/} choice is, in fact, the  {\it correct\/}  one.  It is not clear, from first
principles, how the stability of a sum rule  corresponds to its accuracy.  

The absence of a reliable estimate of the error from choosing $\omega_0$ and 
$\omega$, as well as of the error from truncating the sum over intermediate  states, 
leaves QCD sum rule analyses with systematic uncertainties which are  difficult to 
quantify.  In this respect, they are a lot like lattice gauge theory  calculations  in the
quenched approximation.  Both methods are based, in principle, on QCD,  which  is their
most attractive feature.  However, in their practical  implementation it is  unavoidable
that uncontrolled model dependence emerges.   The result in each  case is  a bit of a
hybrid, a valuable theoretical tool which one must rely on only  with  considerable care.

\section{Quark Models and Related Methods}
\label{Chap2:secFF}

This section discusses quark models and their relatives.  While a QCD analysis  is always 
preferable to a model, there are unfortunately many processes and quantities  of interest 
for which models are the only recourse.  The variety of models, even  commonly used ones, 
is indeed enormous, and there is no hope to survey the field  here.  This  brief section 
explains what is meant by a model, and why a model is distinct from QCD.   Many models are
invented for a very limited purpose, to capture some particular feature  of hadron
phenomenology such as spectroscopy, fragmentation  or weak decay.  Here the focus is  on a
popular model with more general ambitions, the {\it nonrelativistic  quark model.}  This
is  probably the most intuitively accessible model, and it serves  as an excellent
illustrative example.

Consider a $\rho^+$ meson.  In QCD, this state is a complicated collection  of quarks, 
antiquarks and gluons, carrying overall flavor quantum numbers.  Note that  although  a
$\rho^+$ has the flavor of a $u\bar d $ pair, there are in fact many $u$, 
$\bar u$,  $d$ and $\bar d$ quarks in a $\rho^+$, and it is not correct to assign the 
flavor of the  overall $\rho^+$ to any particular ones.  In the nonrelativistic quark 
model, however,  a meson is treated as a bound state of a single quark and antiquark.  
Entirely new  degrees of freedom have been introduced, since these {\it constituent\/} 
$u$ and $d$  quarks are only indirectly related to the quarks of QCD.  They have large 
masses of  order 300 MeV (in contrast to the {\it QCD current\/} quark masses of 
$5-10\,$MeV),  they have small magnetic moments, they are nonrelativistic, they are not 
pair produced,  and they interact with each other through an instantaneous potential.  
This is an  ansatz, not an approximation to or a limit of QCD.

Given these new degrees of freedom, one can then guess a potential and  solve the 
Schr\"odinger equation to find quark wavefunctions.  Magnetic interactions  and other 
effects are introduced as necessary, as perturbations to the nonrelativistic  potential.  
The wavefunctions then may be used to fit or predict physical observables  such as 
spectra, decay constants, or transition rates.  Note that the very idea of a 
nonrelativistic wavefunction is foreign to QCD, so there is no meaningful  sense  in which
the solutions which are obtained are ``correct''.  All that one  can ask is  that the
model be ``predictive'', in that it fit many independent pieces of  data with  few
adjustable parameters.

In principle, models should be constrained to reproduce the known behavior  of QCD in  its
various limits, but this is not always possible.  The chiral limit is a  particular 
problem:  it is difficult to tune the nonrelativistic quark model to obtain  a massless 
pion when  the $u$ and $d$ current masses vanish.  By contrast, heavy quark  symmetry 
provides useful constraints, and it can be used to tune aspects of the  quark model  when
it is applied to bottom and charmed hadrons.

The central problem with models is that it is difficult to accompany their  predictions
with meaningful error estimates.  Since they do not arise as an  expansion of QCD, there
is no small parameter and no systematic corrections to  a controlled limit.  It is very
difficult to guess, when a model is extended  to  a new region, at what level to trust its
predictions.  For example, the  nonrelativistic quark model typically works very well for
hadron spectroscopy,  but this fact gives little insight into its reliability in
predicting form  factors.   It is common practice, unfortunately, to cite uncertainties
due to ``model  dependence''  which are obtained by surveying the predictions of a variety
of models.  This  exercise certainly provides more insight into the tastes of model
builders  than  into the accuracy of their predictions.

Of the wide variety of models currently in use, just a few of  the most  popular ones for
$B$ physics are listed here.  The Isgur-Scora-Grinstein-Wise model is a  version  of the
nonrelativistic quark model which is tuned to the study of semileptonic 
$B$ decays.  The Bauer-Stech-Wirbel model is a quark model on the light front,  used for
weak $B$ and $D$ decays and the exploration of the factorization  hypothesis.  String
fragmentation and flux tube models are used to study heavy quark  fragmentation.  The
Skyrme model, derived from the chiral Lagrangian, is a model of light  baryons  which has
been extended to describe heavy baryons as well.  An alternative  tool  for studying
baryons is the nonrelativistic diquark model.  The ACCMM  (Altarelli  {\it et al.}) model
is used to include initial bound state effects in  inclusive  $B$ decays.  What these
models, and others like them, have in common is that  they  are tuned to specific
particles or specific processes, for which they  typically  work reasonably well.  By
contrast, their predictivity in new contexts is  hard to  assess reliably.  Since it is
unavoidable that models will continue to be an  indispensable tool in $B$ phenomenology,
it is important always to remain  mindful  of their limitations.

\section{Further Reading}
\label{Chap2:secGG}

We have given only the briefest discussion to a  few topics in the theory of
hadronic $B$ physics.  Because of the very general level of the
discussion, references to the original literature have not been included. 
The reader who  wishes to  explore any of these topics further at an introductory level is
invited to consult the many textbooks and reviews which now exist. 

A few examples are:
\begin{itemize}
\item effective Hamiltonians and operator product expansions 
     \begin{itemize}
     \item{\it Dynamics of the Standard Model,} John Donoghue, Eugene Golowich and
     Barry R.~Holstein, Cambridge University Press (1992)
     \item
     G.~Buchalla, A.~J.~Buras and M.~E.~Lautenbacher, 
     {\sl Rev.\ Mod.\ Phys.}\ {\bf 68},\ 1125 (1996)
     \end{itemize}
\item the heavy quark expansion 
     \begin{itemize}
     \item
     M.~Neubert, {\sl Phys.\ Rep.}\ {\bf 245}, 259 (1994)
     \item
     M.~Shifman, in {\it QCD and Beyond,} Proceedings of TASI 95, ed.~D.~Soper, 
     World Scientific (1996)
     \item
     B.~Grinstein, in {\it CP Violation and the Limits of the Standard Model,} 
     Proceedings of TASI 94, ed.~J.~F.~Dono\-ghue, World Scientific (1995) 
     \end{itemize}
\item chiral perturbation theory 
     \begin{itemize}
     \item
     J.~Gasser and H.~Leutwyler, {\sl Phys.\ Rep.}\ {\bf 87}, 77 (1982)
     \end{itemize}
\item lattice gauge theory 
     \begin{itemize}
     \item
     {\it Phenomenology and Lattice QCD,} Proceedings of the Uehling Summer 
     School on Phenome\-no\-logy and Lattice QCD, eds.~G.~Kilcup and S.~Sharpe, World
     Scientific (1995)
     \item
     S.~Sharpe, in {\it CP Violation and the Limits of the Standard Model,} 
     Proceedings of TASI 94, ed.~J.~F.~Donoghue, World Scientific (1995)
     \end{itemize}
\item QCD sum rules
     \begin{itemize}
     \item
     Stephan Narison, {\it QCD Spectral Sum Rules,} World Scientific (1989).
\end{itemize}
\end{itemize}

\acknowledgments

I would like to thank the Stanford Linear Accelerator Center and the BaBar Collaboration
for inviting me to participate in the writing and editing of {\sl The BaBar Physics Book},
and for their hospitality and support during this process.  I am indebted to the
many experts on $B$ physics and others who offered criticism of early drafts, thereby
improving the article and making it more reflective of the diverse points of view of those
working in this field.  Thanks go to Jonathan Bagger, Ikaros Bigi, Robert Fleischer, Fred
Gilman, Benjam\'\i n Grinstein, Dafne Guetta, Laurent Lellouch, Zoltan Ligeti, Michael
Luke, Thomas Mannel, Matthias Neubert, Yossi Nir, Alexey Petrov, Helenka Przysiezniak,
Chris Sachrajda, Nicolai Uraltsev, Mark Wise and Daniel Wyler for their comments and
suggestions.  I am especially grateful to Helen Quinn for her editing of later versions of
the manuscript, as well as for her general wisdom. This work was supported in part by the
United States National Science Foundation under Grant No.~PHY-9404057 and National Young
Investigator Award No.~PHY-9457916, by the United States Department of Energy under
Outstanding Junior Investigator Award No.~DE-FG02-94ER40869, and by the Alfred P.~Sloan
Foundation.  A.~F.\ is a Cottrell Scholar of the Research Corporation.

\end{document}